\documentclass[conference]{IEEEtran}
\IEEEoverridecommandlockouts
\usepackage[space]{cite}
\usepackage{amsmath,amssymb,amsfonts}
\usepackage{algorithmic}
\usepackage{graphicx}
\usepackage{textcomp}
\usepackage{xcolor}
\usepackage{listings}
\usepackage{lstlinebgrd}
\usepackage{booktabs}
\usepackage{url}
\usepackage{tabularx}
\usepackage{multirow}

\begin{document}

\title{Practical Verification of MapReduce Computation Integrity via Partial Re-execution\\
}

\author{\IEEEauthorblockN{Eunjung Yoon}
\IEEEauthorblockA{\textit{Department of Computer Science and Engineering} \\
\textit{Pennsylvania State University}\\
University Park, USA \\
euy109@psu.edu}
\and
\IEEEauthorblockN{Peng Liu}
\IEEEauthorblockA{\textit{College of Information Sciences and Technology} \\
\textit{Pennsylvania State University}\\
University Park, USA \\
pxl20@psu.edu}
}

\maketitle

\begin{abstract}
Big data processing is often outsourced to powerful, but untrusted cloud service providers that provide agile and scalable computing resources to weaker clients.
However, untrusted cloud services do not ensure the integrity of data and computations while clients have no control over the outsourced computation or no means to check the correctness of the execution. 
Despite a growing interest and recent progress in verifiable computation, the existing techniques are still not practical enough for big data processing due to high verification overhead.

In this paper, we present a solution called V-MR (Verifiable MapReduce), which is a framework that verifies the integrity of MapReduce computation outsourced in the untrusted cloud via partial re-execution. V-MR is practically effective and efficient in that (1) it can detect the violation of MapReduce computation integrity and identify the malicious workers involved in the that produced the incorrect computation. (2) it can reduce the overhead of verification via partial re-execution with carefully selected input data and program code using program analysis. 
The experiment results of a prototype of V-MR show that V-MR can verify the integrity of MapReduce computation effectively with small overhead for partial re-execution.
\end{abstract}


\section{Introduction}
\label{sec:intro}
Large-scale distributed computations based on big data processing frameworks such as MapReduce and Apache Spark are often outsourced to public cloud service providers, which use computationally powerful platforms.
While outsourced computation provisions cost-effective and flexible computing resources to a client, the client has no control over the outsourced computation or no means to check the correctness of the execution. The client expects that the cloud provider honestly executes the outsourced program over data and produce the expected output.  

However, a dishonest cloud provider could simply not run a client's code to completion to save on resources or try to cheat and return incorrect computation results to the client. To this end, our goal is to provide guarantees of computation integrity (i.e., verification of execution integrity of the program over the data) that the outsourced computation had not been tampered by adversaries.

Several approaches to verify the integrity of the outsourced computation have been proposed.
We consider the cryptographic and complexity-theoretic approaches \cite{Parno-2013-sp, Rabin-2007-lcs} are still far from practical.
Replication-based result verification approaches \cite{moca-2011, Zhao-2005-p2p, Wang-2013, Wang-2011-cloud} in MapReduce computation rely on redundant computation resources to execute duplicated individual task for all the jobs which incurs high cost. MapReduce computation is often distributed across distributed worker nodes, so the approaches are not scalable.

\begin{figure}[!htb]
\center
\includegraphics[width=0.5\textwidth]{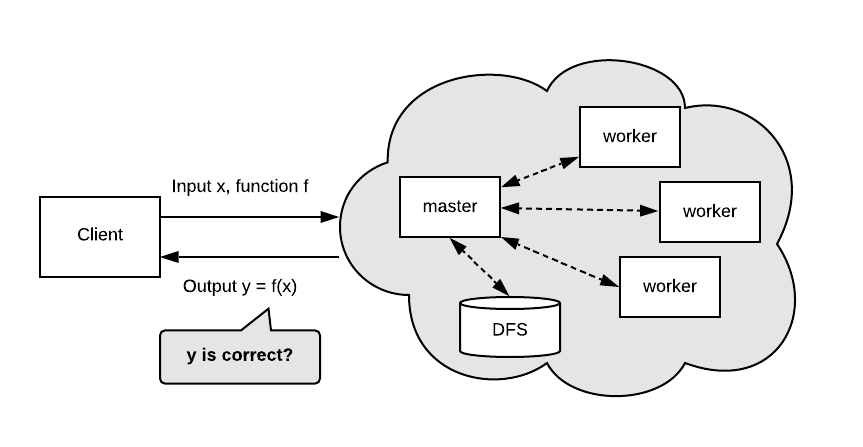}
\caption{Big Data Processing in Untrusted Cloud Environment}
\label{fig:bigdata}
\end{figure}

In this work, we are primarily interested in the computation integrity of the user-defined MapReduce programs which are executed by distributed workers.
MapReduce framework runs arbitrary binary code submitted by a client. The output is determined by the input data, ensuring its final result should be correct as long as the input data remain preserved and the computation is correct (Figure~\ref{fig:bigdata}). 
A naive approach to integrity verification of MapReduce computations is to re-execute every Map and Reduce task, which the workers in the public cloud have executed to verify the correctness of the past computations. However, the overhead of this approach can be very high, particularly if the number of workers is large, as the private cloud is generally less powerful than the public cloud with limited resources. The naive replication-based result-integrity checks in hybrid cloud \cite{Wang-2013} is not practical as well, which is the verification is done by checking the consistency of results among replicas. Most replication-based approaches \cite{Wang-2013, Wang-2011-cloud, Wei-2009-acsac} only provide probabilistic detection of wrong computation. Large scale data processing often requires a large number of replicas that incur very high overhead.
We argue that a better approach is to selectively and partially re-execute tasks on the trusted verifier. 

The goal of this work is to provide integrity guarantees of outsourced computations to the clients by efficient computation integrity auditing of workers. In particular, given the input data, the execution of the program over the data can be checked by detecting tampered computation through partial re-execution, which is the practical verification approach with smaller overhead compared to naive full re-execution or replication approach.

To this end, we propose \emph{Verifiable MapReduce} (V-MR), a solution to detect tampered MapReduce computation and identify the malicious workers that produced the wrong output via \emph{partial re-execution}. V-MR collects execution traces during the original execution and preprocesses the traces using data and control flow information to find data slice and program slice (executable slice) for partial re-execution. V-MR performs the partial re-execution of Map and Reduce tasks with selected input data (data slice) and program code (program slice) for efficient auditing the computation results and therefore detecting any malicious workers that did not execute the client's program as expected. 
\\

To summarize, we make the following contributions.
\begin{itemize}
\item
We introduce the efficient and practical Verifiable MapReduce (V-MR) for verifying the integrity of the MapReduce computations via partial re-execution with the reduced re-execution overhead.

\item
We introduce input data slicing and program slicing by control flow filtering based on static and dynamic analysis to reduce the input dataset which is used for re-execution.

\item
We present a new program transformation technique using the bytecode engineering library, which automatically transforms the original program to the new code to re-execute for the verification of computation integrity.

\item
We show the experimental results with several MapReduce applications, the reduced overhead of V-MR compared to naive replication approach that requires full re-execution or random input selection. 
\end{itemize}

\section{Related Work}
\label{sec:related}
\textbf{Replication-based Verification:}
SecureMR \cite{Wei-2009-acsac} adopts a decentralized replication-based integrity verification scheme, and utilizes the
existing architecture of MapReduce for replication purposes.
Replication is broadly used in distributed systems to improve reliability and fault-tolerance, as well as for computation integrity in big data computing such as MapReduce. However, the assumption that there is no collusion between workers is needed when the result checking is based on voting by the workers. Furthermore, redundant and replication based approach incurs high cost. Our solution does not require the assumption and reduces the overhead of the replication. 

\textbf{Hardware-based Verification:}
Recently, several approaches have been proposed to protect computation integrity in the cloud leveraging the hardware protection of Intel SGX, which recently has been receiving much attention.
For example, Haven \cite{Baumann-2014} runs service instances in
hardware-protected SGX enclaves \cite{McKeen:2016} and attest to a hash of the code so that the instance owner can verify that it was launched correctly. VC3 \cite{Schuster-2015}, M2R \cite{Dinh-2015}, and Ohrimenko et al. \cite{Ohrimenko-2015} use SGX to perform data analytics, MapReduce computations,
and machine learning computations while ensuring
confidentiality and integrity. This ensures the integrity of the software stack loaded, but it does not gurantee the execution integrity of the program atop the software stack. Haven \cite{Baumann-2014} relies on Intel SGX for shielded execution of unmodified legacy applications to protect the confidentiality and integrity of a program and its data from the platform on which it runs.

\textbf{Proof-based Verification:}
Proof-based verifiable computation is a one of the approaches that verifies the integrity of the remote computation. In this class of works, the verifier (the private cloud in our setting) will generate a set of constraints C about the runtime values of variables in the program. During the execution, the prover (the public cloud in our setting) will generate the proof of the computation based on the runtime value of variables. The verifier then will perform a set of tests on the proof based on C. If the program was executed faithfully, the test will pass; otherwise, the test will fail, except for a very small probability. Systems following this direction include Pepper \cite{Braun:2013}, Buffet \cite{wahby2015efficient}, TinyRAM \cite{Sasson-2013}, etc. These approaches incur an expensive overhead in both the verifier and the prover amd can only support simple programs.

\textbf{Record-replay approaches:}
VM replay ~\cite{Bressoud-1995}, ReVirt ~\cite{dunlap-2002-sigops}, Ripley ~\cite{kvikram2009ripley}, POIROT ~\cite{Kim2012-SEC}, and OROCHI ~\cite{Tan2017-SOSP} are based on record-and-replay techniques. In particular, POIROT and OROCHI's approaches are mostly close to our approach but their works are for web applications while our work is for MapReduce applications.

\section{Threat Model}
The attack surface we consider in this work is that of MapReduce applications which
run on distributed worker nodes. Malicious adversaries have capability to manipulate computation results
or the user-defined map and reduce functions to produce incorrect results.

We assume that a client’s program (user-defined map and reduce function) is benign
when she submitted the job request to the cluster, in which the client is honest who
will never intentionally try to compromise the worker nodes. However, a client’s virtual
machine (or container) images in the public cloud are not trusted, which execute the
user-defined program. The images may have security vulnerabilities that an adversary
can exploit. A powerful adversary can alter input, intermediate and final outputs of
MapReduce computations or tamper the user-defined program to produce incorrect results.

We assume the master node is trusted, backed by strong security, which is responsible 
for job scheduling and resource management of worker nodes in the cluster while workers
are not trusted. We also assume the Distributed File System (DFS) is trusted, which
manages the file system namespace and regulates access to files by clients.

The verifier is assumed to be trustworthy and protected from attack. The verifier
collects the execution trace logs provided by the workers in the public cloud and the
$<input, output>$ of each task that has been executed and checks if worker nodes have
executed the computation as expected.

An important assumption for our re-execution based verification approach is that the
MapReduce program (i.e., map and reduce function) is deterministic. The assumption
guarantees that all executions of the program with the same input always produce the
same output. Thus, when given the same input, an honest worker always returns the
correct result for its computation task while a malicious worker may return arbitrary or
manipulated results. 

\subsection{Attacks on the MR computation}
In this work, we consider two types of attacks on MR computation.

\textbf{Cheating Computation:}
A cheating worker tries to skip some part or the entire of the computation and return only a subset of results to reduce its computing cost.  Most machine learning and data mining applications involve iterative computations over large datasets so skipping the execution of part of the loops in the program (that has the operation of writing the output to the local disk or DFS) could be an attack example. Cheating workers are relatively easy to detect as discussed in \cite{YoonS14-ccgrid}.

\textbf{Malicious Computation:}
A malicious worker can disrupt the computation of map or reduce instances by tampering with the code or data submitted by a client, resulting to yield the incorrect computation results. Even one worker's incorrect computation can yield the overall incorrect final results. In this work, we focused on malicious computation attack.

\subsection{Integrity of Execution Trace Log}
In this work, the execution trace log recorded during normal execution is essential for auditing the computation results.
We assume that the execution trace logs are protected from the integrity attack in which an adversary cannot alter the logs, while an untrusted worker could return the wrong computation results. 
We assume the hypervisor is trusted and leverage virtual machine based techniques \cite{dunlap-2002-sigops} that provide hypervisor-supported inspection or tamper-evident logging \cite{Crosby:2009, Ma:2009} for prot{}ecting the audit logs from the integrity attack.
We make the audit logs are append-only such as LogFAS \cite{Yavuz:2012}. LogFAS is a PKC-based secure audit logging scheme that can produce publicly verifiable forward-secure and append-only signatures without requiring any online trusted server support.

\section{Overview}
\label{sec_overview}
Our focus is on verifying the integrity of MapReduce computation in untrusted cloud computing environment. 
Our approach is to use \emph{partial re-execution} for verifying whether untrusted MapReduce workers provided correct computation results for assigned program and data, by re-executing only a part of the original MapReduce computations to minimize the re-execution overhead.

Through the experiment with multiple MapReduce applications, we observed that there is much similarity among the execution traces of workers as workers assigend for the job execute the same MapReduce program in parallel. 
Each Map and Reduce task execution follow a relatively small number of execution paths (control flow) and the difference of execution paths is mostly caused by different input data.
Our approach is similar to record and replay approaches \cite{dunlap-2002-sigops,Bressoud-1995,kvikram2009ripley} which record computation history (execution traces) and replay (re-execution). 

Our approach comprises of two stages: online tracing and offline verification. Online tracing is to record execution traces of the user-defined MapReduce application on every worker node during normal execution.
The offline verifier analyzes the execution traces to identify and select the requests to re-execute and prepare the new code (if optimization is possible) and the input data for partial re-execution.
The verifier re-executes the chosen requests with chosen input data and compares the outputs of the original execution and re-execution. If the outputs are different, the verifier rejects the computation for the job and reports the malicious worker that executed the request and produced the corresponding incorrect output.

\begin{figure*}[ht!]
\centerline{\includegraphics[width=0.9\textwidth]{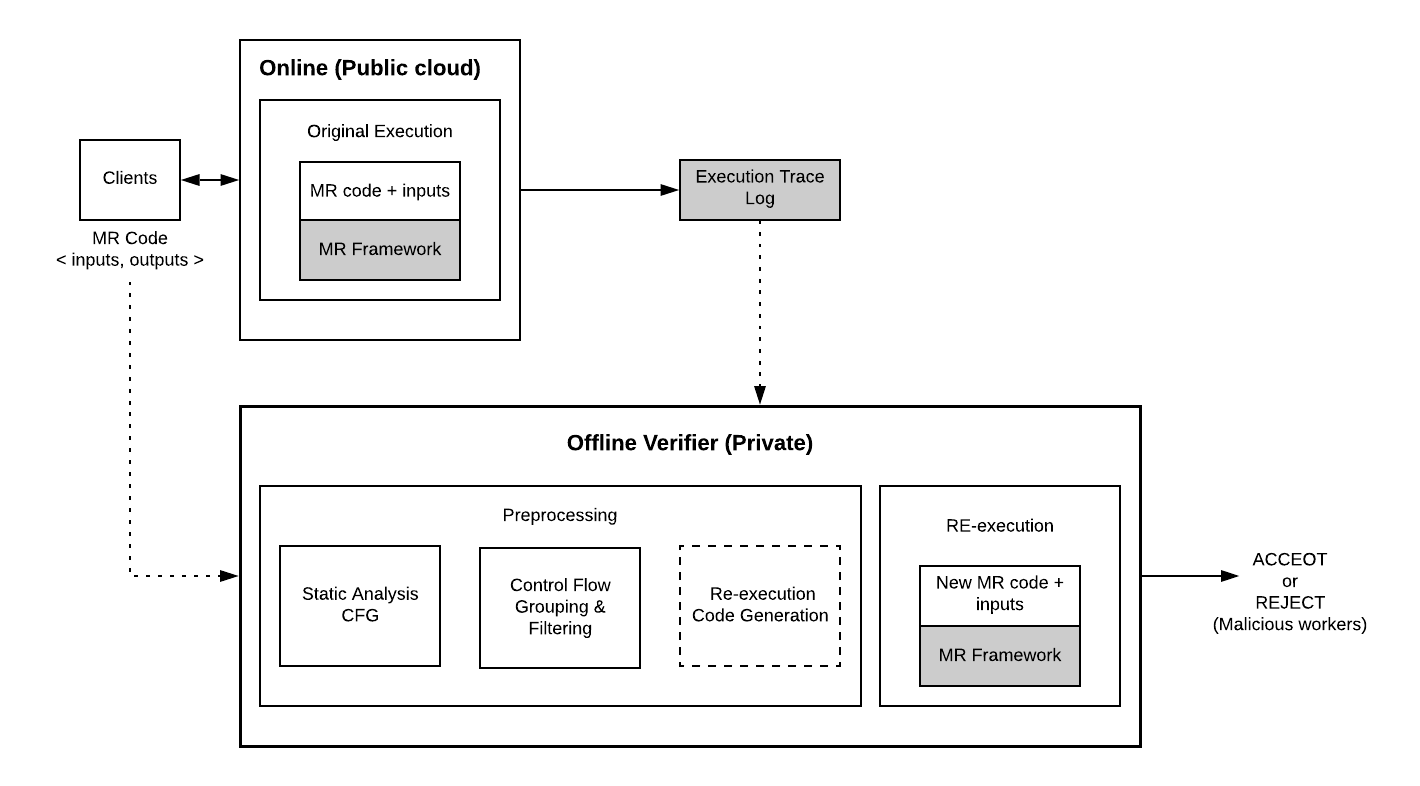}}
\caption{Overview of V-MR design}
\label{fig:design}
\end{figure*}

The architecture of V-MR is illustrated in Figure~\ref{fig:design}.
We consider parallel and distributed applications such as MapReduce applications run on worker nodes (computing nodes) in a cluster. 


A client submits a job with the user-defined MapReduce program and data to the cluster and receives the computation results.
All MapReduce workers assigned for a job are asked to submit the execution traces of the program as proof of the correct execution, which is recorded during program execution. The offline verifier checks if all workers performed the submitted job correctly as expected using the execution traces.



V-MR comprises of 1) offline preprocessing of bytecode-level execution traces, 2) selection of requests for the re-execution based on control flow and data flow analysis, re-execution code generation if duplicate computations across workers are discovered, and 3) re-execution of the original or new code that has been generated, and 4) the comparison of the outputs from the two executions and make a decision to accept or reject the computation result. For the rejected computation results, V-MR identifies the corresponding workers that computed the requests and marks those as malicious workers.

\subsection{Static Control-Flow Analysis}
To reduce the overhead of verification, we use static analysis and generate the Control Flow Graph (CFG) from Java bytecode of the user-defined MapReduce program. Static analysis helps reduce the overhead of dynamic analysis and explores all execution paths without running the program, typically from source code. Control flow analysis determines the execution order of program instructions. It brings a global overview of the execution paths that the input data can take during the execution process. In this context, we represent a Java method of a class with its CFG that consists of a set of its execution paths because we use the MapReduce framework in Java.

V-MR generates control flow graphs for map and reduce function using static analysis of the user-defined MapReduce program. 
Control Flow Graph G=(V,E) is a directed graph which represents the execution flow of the program. Where V is the set of vertices representing basic blocks, and E is the set of edges - an edge from basic block $BB_{1}$ to basic block $BB_{2}$ ($BB_{1}, BB_{2}$ $\in$ V) indicates that execution of block $BB_{2}$ can immediately follow the execution of block $BB_{1}$. A basic block is the sequence of executable instruction with a single entry point at the beginning and exit point at the end.
Control flow analysis consists in computing the control flow graph of a method, and in performing analyses on this graph. 

The Control Flow Graph is used in the verification phase for detecting invalid control flow paths and filtering out the corresponding requests from the verification, which followed the invalid execution path.

\begin{figure}[!htb]
\centering
\includegraphics[width=0.5\textwidth]{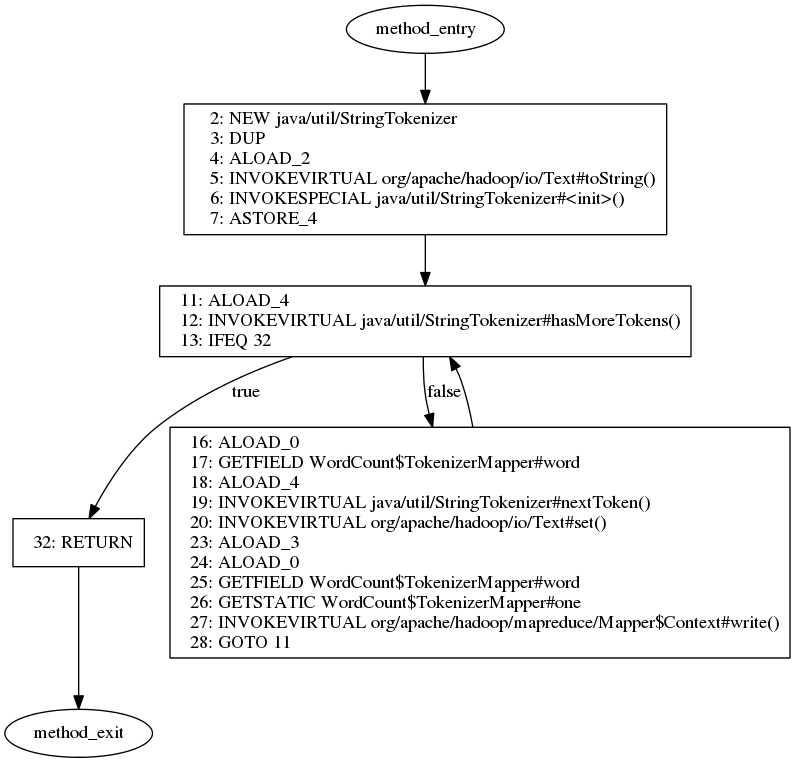}
\caption{Control Flow Graph of WordCount map()}
\label{fig:wordcount_map_cfg}
\end{figure}

\begin{figure}[!htb]
\centering
\includegraphics[width=0.5\textwidth]{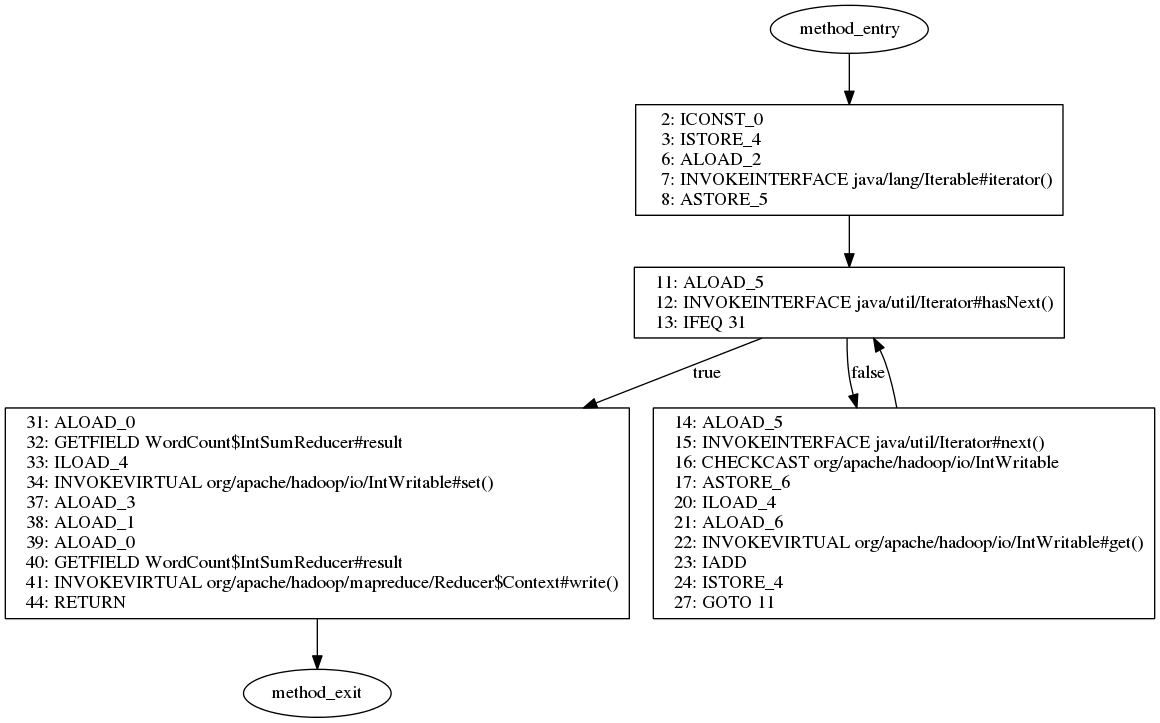}
\caption{Control Flow Graph of WordCount reduce()}
\label{fig:wordcount_reduce_cfg}
\end{figure}

A client submits the job to the cluster with the compiled MapReduce program in binary form (jar file). We use the ASM bytecode engineering library to generates the CFG from the bytecode of the program. Figure \ref{fig:wordcount_map_cfg} and Figure \ref{fig:wordcount_reduce_cfg} show the generated and visualized CFGs for map and reduce function of WordCount program using the ASM and GraphViz~\cite{GraphViz:2002}.
V-MR uses the control flow graph in the verification phase for detecting invalid execution paths and filtering out the corresponding requests from the verification, which followed the invalid execution path. 

\definecolor{dkgreen}{rgb}{0,0.6,0}
\definecolor{gray}{rgb}{0.5,0.5,0.5}
\definecolor{mauve}{rgb}{0.58,0,0.82}

\lstset{frame=tb,
  language=Java,
  aboveskip=3mm,
  belowskip=3mm,
  showstringspaces=false,
  columns=flexible,
  basicstyle={\small\ttfamily},
  numbers=left,
  numberstyle=\tiny\color{black},
  keywordstyle=\color{blue},
  commentstyle=\color{dkgreen},
  stringstyle=\color{mauve},
  breaklines=true,
  breakatwhitespace=true,
  tabsize=3,
}


\begin{figure}[htb!]
\center
\begin{minipage}{.4\textwidth}
\begin{lstlisting}[
    basicstyle=\fontsize{8}{11}\ttfamily,
]
    public void map(Object key, Text value, Context context
                    ) throws IOException, InterruptedException {
      StringTokenizer itr = new StringTokenizer(value.toString());
      while (itr.hasMoreTokens()) {
        word.set(itr.nextToken());
        context.write(word, one);
      }
    }
    ....

    public void reduce(Text key, Iterable<IntWritable> values,
                       Context context
                       ) throws IOException, InterruptedException {
      int sum = 0;
      for (IntWritable val : values) {
        sum += val.get();
      }
      result.set(sum);
      context.write(key, result);
    }
\end{lstlisting}
\end{minipage}
\caption{WordCount example code snippet}
\label{fig:wordcount}
\end{figure}

\subsection{Tracing during Original Execution}
V-MR records the execution trace of the user-defined MapReduce program during normal execution, which runs parallel on multiple worker nodes. 
An execution trace is a sequence of instructions which represents the execution history of a program and provides the details of a program's dynamic behavior. 
V-MR collects and analyzes the execution traces to select the partial input data and code for efficient re-execution.

Recording all of the execution would not be practical due to significant time and space overhead.
To reduce the tracing overhead, V-MR logs the selected execution trace in terms of control flow (\emph{control flow trace}), as not all of the trace data is of our interest.
V-MR also logs input $<key, value>$ pair, branch condition and count if it is a loop, output $<key, value>$ pair for map() and reduce() function.

V-MR performs partial re-execution during the verification phase. For the verification, V-MR re-executes the program with the same input that processed the corresponding request.

\definecolor{dkgreen}{rgb}{0,0.6,0}
\definecolor{gray}{rgb}{0.5,0.5,0.5}
\definecolor{mauve}{rgb}{0.58,0,0.82}

\lstset{frame=tb,
  language=Java,
  aboveskip=3mm,
  belowskip=3mm,
  showstringspaces=false,
  columns=flexible,
  basicstyle={\small\ttfamily},
  numbers=left,
  numberstyle=\tiny\color{gray},
  keywordstyle=\color{blue},
  commentstyle=\color{dkgreen},
  stringstyle=\color{mauve},
  breaklines=true,
  breakatwhitespace=true,
  tabsize=3,
  moredelim=**[is][\color{red}]{@}{@},
}


\begin{figure}[]
\center
\begin{minipage}{.4\textwidth}
\begin{lstlisting}[
    basicstyle=\fontsize{8}{11}\ttfamily,
]
    public void map(Object key, Text value, Context context
                    ) throws IOException, InterruptedException {
      @mapID = getMapId()@              
      @in = (<key, value>)@    

      StringTokenizer itr = new StringTokenizer(value.toString());
      @L0@      //Basic Block
      while (itr.hasMoreTokens()) 
      @bcondition = itr.hasMoreTokens())@
      {
      @L1@      
        word.set(itr.nextToken());
        context.write(word, one);
        @out = (<word, one>)@
      }
      @L2@
    }
    
\end{lstlisting}
\end{minipage}
\caption{Instrumented code snippet for WordCount map() function. {\small The figure shows the bytecode instrumentation at the source level for clarity, but the actual
instrumentation is done directly at the bytecode level. The statements in red are those instrumented.}}
\label{fig:wordcount_instrument}

\end{figure}

\subsubsection{Tracing Control Flow}
During the original execution, V-MR records the \emph{control flow trace} of the application for each request which each map() or reduce() instance processes. 
A control flow trace captures the complete path followed by a program during the execution at the level of Basic Block. However, tracing the entire execution of the program incurs a high space and time overhead so we trace the bytecode instructions that caused the control flow transfer, such as conditional branch instructions (e.g., ifeq, ifne, if\_icmpeq) and unconditional branch instructions (e.g., goto, jsr, ret). 
V-MR also records the requests that executed the basic block that includes the calls to context.write() which contributed to the output of the request.

A MapReduce job is assigned and executed by many workers in parallel with the same user-defined program so the computation performed by each Map or Reduce task is expected to follow the same control flow given the same input data. In terms of execution traces, the sequence of bytecode instructions executed is similar (not equal) but different control-flow paths by input-dependent control transfers. 

In WordCount example (\ref{fig:wordcount} and \ref{fig:wordcount_map_cfg}), the control flow traces of map() instances with different input records are as follows. The number in the control flow trace represents the basic block index. To minimize the overhead of tracing, V-MR traces the control flow transfer only. The execution of $BB_1$ is not recorded as the basic block doesn't have control flow transfer.  All mappers commonly execute $BB_1$ for all requests. During the preprocessing phase, V-MR groups the requests using control flow traces. \\

{\tt
\small{
Input record <1, "test input">:

CTF: 2, 3, 2, 3, 2, 4 \\

Input record <1, "test">:

CFT: 2, 3, 2, 4
}
} \\

\begin{figure}[htb!]
\center
\begin{minipage}{.45\textwidth}
\begin{lstlisting}
13: aload         4
15: invokevirtual #8        
18: ifeq          47
........................
21: aload_0
22: getfield      #4       
25: aload         4
27: invokevirtual #9       
30: invokevirtual #10      
33: aload_3
34: aload_0
35: getfield      #4       
38: getstatic     #11      
41: invokevirtual #12       
44: goto          13
.........................
47: return
\end{lstlisting}
\end{minipage}
\caption{Java bytecode instructions for lines in 4-7 (while loop) in Figure \ref{fig:wordcount}}
\label{fig:wc_bytecode}
\end{figure}

\subsection{Partial Re-execution}
The goal of our work is to produce the practical verification of result integrity via efficient partial re-execution. 
The re-execution cost needs to be minimized for efficient verification while the correctness and completeness of the verification need to be guaranteed as well. Finding the solution that satisfies both conditions is challenging.

Through our experiments with several MapReduce applications, we observed that there are a relatively small number of control flow paths among the executions of MapReduce instances across workers. That is because the distributed workers execute the same user-defined MapReduce program in parallel on their assigned data partitions for a given job. Different inputs may cause different control flows yet a relatively small number of execution paths (i.e., control flow paths) per each Map or Reduce task.  
To achieve efficient integrity verification using partial re-execution, we aim to reduce the input data set and the instructions of the re-execution code.
Based on our observation, we leverage the control flow trace of each Map and Reduce instance for selecting the part of the input data set (\emph{input data slice}). 

V-MR groups the Map and Reduce instances that have the same control flow into a control flow group.
Then, V-MR identifies the control flow group that did not lead to the outputs by checking whether the control flow trace of the group shows the execution of the basic block that has the instruction to emit the output (calls to context.write() in Hadoop v2).

\subsubsection{Input Data Slicing for Partial Re-execution}
Naively re-executing all instances of Map and Reduce function on all input data for the verification incurs the high overhead. To amortize the verification cost, V-MR re-executes only parts of the MapReduce function call instances after it filters out any input records that did not contribute to the output.
MapReduce applications have a relatively simple control flows due to the defined programming model of the map and reduce function. 
Based on our observation that a MapReduce program has multiple identical computations across workers, we use the control flow traces for selecting parts of input data (input data slice) for partial re-execution, similar to POIROT's approach  \cite{Kim2012-SEC}.  
V-MR groups the Map and Reduce instances that have the same control flow trace into a control flow group such that all instances in a control flow group have the same control flow trace.

Grouping instances that have the same control flow helps to locate the deviated control flow group, which the verifier marks the workers in the group as suspicious.
All Map and Reduce function call instances in a control flow group have the same control flow trace. V-MR selects the instances to be re-executed in each control flow group. If V-MR identifies any control flow group that did not take eligible execution paths, that is, one of  the all execution paths found in the static CFG, V-MR rejects the computation results. In this step, V-MR can optionally select an aribitrary instance from the control flow group and re-execute it before it stops the entire verification. 

Control flow based filtering helps to reduce the input data set used for the verification. However, we need a better approach for more sophisticated attack that does not change the control flow of the original program to avoid the detection while producing tampered computation results.

\subsubsection{Program Slicing for Partial Re-execution}
\label{subsec:instructions}
MapReduce applications have relatively simple control flows due to the defined programming model of the map and reduce function. The control flow-based filtering of the Map and the Reduce instances using their control flow trace and static CFG enables the detection of diverted control flow paths. However, control flow-based filtering may not be sufficient for detecting a more sophisticated attack that does not change the control flow. 

For the correctness and completeness of integrity verification, we also consider data dependency of the program. To minimize the re-execution overhead, we try to minimize the program instructions to re-execute as well as the input data set. V-MR identifies the instructions in a control flow group that do not change the output of the program across all instances and eliminates those instructions when it generates the re-execution code using the program transformation. 
However, V-MR needs to re-execute any instructions that produce different outputs depending on different inputs. To identify those instructions that need to be included in the re-execution code, we adopt the taint-based dependency analysis \cite{Newsome05dynamictaint}. Similar to POIROT's approach, V-MR selects an arbitrary instance from a control flow group and re-executes the instance with an input record computed on during original execution. V-MR performs taint-based dependency analysis at the level of bytecode.

\begin{figure}[!hb]
\hspace*{-0.5in}
\centering
\includegraphics[width=0.6\textwidth]{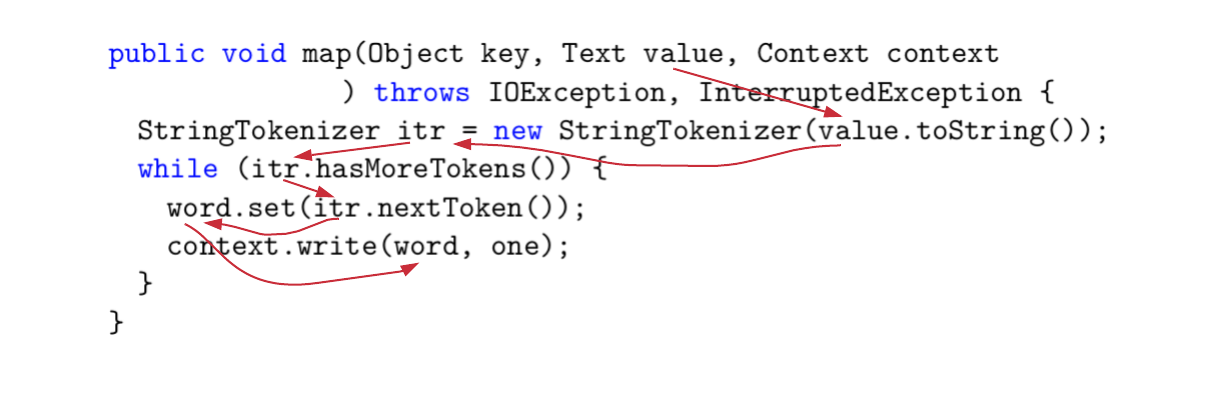}
\caption{Data Dependency Tracking with Taint analysis (WordCount)}
\label{fig:dataflow_wc}
\end{figure}

An instance of Map or Reduce function emits the output only by calling \texttt{context.write()} method (in Hadoop V2) so we consider the calls to \texttt{context.write()} and an input record $<key, value>$ pair (i.e., parameters of the function) that leads to the method call. V-MR uses taint-analysis for tracking the data dependency but from one arbitrary selected instances to reduce the tracing overhead. During the re-execution of the selected instance, V-MR marks these input parameters (variables) as "tainted" upon the start of the map or reduce function. During the execution of the instance, V-MR marks the instructions that read the tainted variables and their outputs tainted as well (Figure~\ref{fig:dataflow_wc}). These tainted instructions must be included in the re-execution code as they produce different outputs for different inputs while taking the same control flow path.

The instructions that depend on the tainted variables cannot be eliminated and must be included in the re-execuion code.
If there are identical computations, in which instructions that do not depend on the tainted variables and control flow instructions, the instructions can be eliminated when the re-execution code is generated.

Taint-based dependency tracking produces the following taint set for the execution of each instruction of the code in Figure \ref{fig:dataflow_wc}. \\

{\tt
\small{
Taint set:\\
\indent 1  \{$value$\} \\
\indent 2  \{$itr, value$\} \\
\indent 3  \{$itr, value$\} \\
\indent 4  \{$word, itr, value$\} \\
\indent 5  \{$word, itr, value$\} \\
}
}

For example, the following instances produce the same control flow trace and fall into the same control flow group. While the two different executions follow the same control flow path, their inputs and outputs are different. \\

{\tt
\small{
Original WordCount Code (Figure \ref{fig:wordcount}): 

Input <1, "test"> 

CFT: 2, 3, 2, 4 

Output: <"test", 1> \\

Attacker's WordCount Code (Figure \ref{fig:wordcount_attack}): 

Input <1, "attack"> 

CFT: 2, 3, 2, 4 

Output: <" ", 1> \\
}
}

In this case, control flow analysis solely cannot detect the attack as the two executions follow the same control flow path (2, 3, 2, 4) so the re-execution is needed.  We use the taint-based data dependency analysis to determines which instructions need to be included in the re-execution code. In this control flow group, let's suppose that the first instance with the input $<1, "test">$ is selected for taint analysis,  then the instructions 3, 4, 5, 6 must be included (no elimination of the instruction). When V-MR re-executes the second instance (attacker's code) with the selected instructions 3, 4, 5, 6 on the input record $<1, "attack">$, the output is $<"attack", 1>$, which is inconsistent with the actual output $<" ", 1>$ from the original execution. V-MR detects the inconsistencies between the outputs and rejects the computation result. V-MR identifies the malicious worker that emitted the wrong output.

\subsubsection{Program Transformation for Generating Re-execution Code}
V-MR transforms the original program into a new program that will be used for the re-execution of selected Map and Reduce instances using the ASM bytecode manipulation library. The ASM reads the original bytecodes (the application class files) and manipulates the bytecodes to generate executable class files that will be used for partial re-execution. In this work, the ASM removes the bytecode instructions identified by our optimization in the previous step (\ref{subsec:instructions}) from the original program. 

A program transformation should preserve the semantic properties of the original program which should not break the semantics of the target program or cause any difference in outputs. The program syntax of the two programs is different but semantically equivalent, which yields the same output from the computation (execution of the program) with the same input.
A program transformation is a partial function, mapping a program $P$ to another program $P'$. 
We use the ASM bytecode engineering library to transform the original MapReduce program $P$ into a new MapReduce program $P'$ which the program slicing has been applied to $P$.
V-MR re-executes the selected instances with the transformed MapReduce program during the verification phase.

\subsubsection{Re-execution}
Once V-MR completes the selection of Map and Reduce instances to re-execute and the generation of re-execution code during the preprocessing phase, V-MR partially re-executes the instances on the re-execution code and selected input records for the verification of the computation integrity. 
The verifier runs the re-execution code on the MapReduce runtime in the private cluster/private cloud. 

V-MR partially re-executes the selected instances in each control flow group on the re-execution code and chosen input records. It then compares the outputs with those of the original execution to check whether the computations in each control flow group are correct.
If the outputs of the re-execution are the same as the outputs of the original execution, V-MR accepts the original computation as correct. 
If the outputs from two executions are different, V-MR rejects the computation as incorrect and identifies the workers that produced the wrong computation results.
V-MR checks the computation integrity of the map and the reduce function separately. If and only if the computation integrity of all mappers is verified and accepted, V-MR proceeds to the next step to verify the computation integrity of the reducers.

\section{Implementation}
\label{V-MR_impl}

In this section, we describe our prototype implmentation of a prototype of V-MR for MapReduce applications. V-MR's components are implemented in Python and the ASM bytecode library.

\subsection{Bytecode Instrumentation}
There are two types of bytecode instrumentation techniques -  the static instrumentation and the dynamic instrumentation. The static instrumentation inserts all instrumentation code before the program starts execution, while dynamic instrumentation injects the instrumentation code into the running process at runtime. 
In this work, we use the dynamic instrumentation to check the behavior of the program at runtime.

We use the ASM bytecode engineering library to instrument the bytecodes of the MapReduce applications. Our selection of the ASM bytecode engineering library aside from other instrumentation libraries is for better performance and a small memory footprint.
V-MR dynamically instruments the MapReduce application classes loaded into the JVM through a Java agent.  Our Java agent is built on the ASM bytecode manipulation framework with the event-based ASM API, which contains a pre-main method that implements a class transformer (\texttt{ClassFileTransformer}) that is composed of a class reader (\texttt{ClassReader}) and a class writer (\texttt{ClassWriter}). The Java agent records the control flow traces of the target program during normal execution without any modification of the program. The control flow trace logs from workers are aggregated and sent to the verifier (Figure~\ref{fig:java_agent}).

\begin{figure}[htb!]
\center
\hspace*{0.5cm}
\begin{minipage}{.45\textwidth}
\begin{lstlisting}[
    basicstyle=\fontsize{8}{11}\ttfamily,
]
package asm;
import java.lang.instrument.Instrumentation;
public class Agent {
    public static void premain(
        String args, Instrumentation inst) {
        inst.addTransformer(new CustomClassTranformer());
    }
}
...

public class CustomClassTransformer
    implements ClassFileTransformer {
    public byte[] transform(...) {
        ClassWriter cw = new ClassWriter(true);
        ProfilerVisitor cv = new ProfilerVisitor(cw);
        ClassReader cr = new ClassReader(classfileBuffer);
        cr.accept(cv, false);
        return cw.toByteArray();
    }
}
\end{lstlisting}
\end{minipage}
\caption{ASM Java Agent code snippet}
\label{fig:java_agent}
\end{figure}

\subsection{Re-execution Code Generation}
V-MR generates efficient re-execution code using program slicing \cite{Weiser:1981} which extracts an executable subset (a slice) of the original program. 
We use the ASM library with the template which we created from the taint-based dependency tracking. For each control flow group, the new execution code is generated using ASM instrumentation with the common bytecode instructions extracted from the original code, which are also Java bytecodes (executable class file). V-MR runs the new code with the same input data for selected worker node to be checked and compare the results of two executions for the computation integrity.

\section{Evaluation}
\label{sec_evaluation}
In this section, we show the effectiveness and the performance of our approach. First, we used three attack scenarios to show the effectiveness of V-MR. Second, we show the experiments results that we performed and the metrics that we used to evaluate the performance of V-MR.
Table~\ref{tab:HadoopApp} describes the MapReduce applications we used for the experiment. We show the tracing overhead of V-MR and running time of partial re-execution compared with naive full re-execution.

\subsection{Experimental Setup}
In this section, we describe the experimental setup for evaluating a prototype of V-MR. We built an experiment Hadoop cluster with two physical machines of the 8-core CPU of 3.40 GHz. The experiment cluster consists of 6 virtual machines (VM), which are provisioned from these two physical machines using VirtualBox and have Ubuntu 14.04 LTS and Hadoop 2.6.2 installed. Among these VM nodes, one VM is configured as a master node and the other VMs configured as worker nodes. We also built a separate private verifier cluster with one physical machine of the 8-core CPU of 3.40 GHz, which consists of 4 virtual machines that have Ubuntu 14.04 LTS and Hadoop 2.6.2 are installed.

Hadoop consists of two layers: MapReduce for data processing, and Hadoop Distributed File System (HDFS) for data storage. A Hadoop cluster includes a single master node and multiple slave nodes. The master node acts as bot the ResourceManager for scheduling map or reduce jobs and the NameNode for hosting HDFS indexes. Each slave node acts as both the NodeManager for conducting the map or reduce operations and the DataNode for storing data blocks in HDFS.

\begin{table}[t]
\caption{Description of MapReduce applications} 
\label{tab:HadoopApp}
\centering
\begin{tabular}{|p{4cm}|p{4cm}|}
\hline
\textbf{Program}       & \textbf{Description}  \\ 
\hline   
Word Count    & A MapReduce program that reads text files and counts how often words occur.  \\ \hline
Inverted Index  & A MapReduce program that is an index data structure storing a mapping from content, such as words or numbers, to its locations in a database file, or in a document or a set of documents.    \\ \hline
Web log analysis (Hit Count) & A MapReduce program that analyze Web application logs such as the hit count and the frequency of URLs. \\ \hline
Web log analysis (Frequency) & A MapReduce program that makes frequency distribution of the number of hits received by each URL sorted in ascending order, which simply sorts the list based on the number of hits that is calculated by the above Web log analysis (Hit Count) program. \\ \hline
\end{tabular}
\end{table}

\subsection{Attack simulation}
In this section, we use several attack cases to show that our verifier is effective to detect the tampered MapReduce computations.

To simulate our attack scenarios, a malicious code may be injected using the Hadoop's ``Zip Slip'' vulnerability (CVE-2018-8009) that has been recently found.
The vulnerability is exploited using a specially crafted archive that holds directory traversal filenames (e.g. \texttt{ ../../evil.sh}). The Zip Slip vulnerability \cite{ZipSlip} can affect numerous archive formats, including tar, jar, war, cpio,  apk, rar and 7z. 
Zip Slip is a directory traversal vulnerability that can be exploited by extracting files from an archive. An attacker can gain access to parts of the file system outside of the target folder in which they should reside. The attacker can then overwrite executable files, thus achieving remote command execution on the victim's machine. The attacker also can overwrite configuration files or other sensitive resources. The vulnerability can be exploited on both client machines and servers.

\emph{Distributed Cache} is a facility provided by the Hadoop MapReduce framework. It caches files when needed by the applications. It can cache read-only text files, archives, jar files, etc. For example, a user-defined MapReduce program will be available on each data node where MapReduce tasks are running.
Thus, we can access files from all the data nodes on the MapReduce job.

To simulate the attack, Hadoop's \emph{Distributed Cache} facility is used.
The attacker's code is copied to File System and then set up the application's JobConf to add the files to \emph{Distributed Cache} and modify the Mapper or Reducer's code to use the cached files (i.e., attacker's code).

\subsection{Effectiveness}
Our primary focus is on detecting integrity attacks to MapReduce computation where a compromised worker produces the incorrect outputs. We show the effectiveness of our approach with following attack scenarios.

\subsubsection{Attack Scenario 1}
The attacker's MapReduce code ran on one of the worker nodes and produced the incorrect output.
Figure \ref{fig:wordcount_attack} shows the attacker's WordCount code snippet. The attacker's code can exploit the integrity attack to WordCount application, which calls method \texttt{foo()} in the map function. This code disrupts the computation output by modifying the word to be emitted to an empty string right before emitting the output, which results in the output with words of the empty string. Figure \ref{fig:wordcount_attack_bytecode} shows the corresponding bytecodes of map function which invoke the attacker's method call, \texttt{foo()}.


\begin{figure}[htb!]
\hspace*{0.5cm}
\begin{minipage}{.45\textwidth}
\begin{lstlisting}[    
		basicstyle=\fontsize{8}{11}\ttfamily,
        linebackgroundcolor={%
        \ifnum\value{lstnumber}>15
            \ifnum\value{lstnumber}<18
                \color{red!20}
            \fi
        \fi
        \ifnum\value{lstnumber}>21
            \ifnum\value{lstnumber}<25
                \color{red!20}
            \fi
        \fi},
]

public class WordCount {

    public static class TokenizerMapper
        extends Mapper<Object, Text, Text, IntWritable>{

    private final static IntWritable one = new IntWritable(1);
    private Text word = new Text();

    public void map(Object key, Text value, Context context
                    ) throws IOException, InterruptedException {
        StringTokenizer itr = new StringTokenizer(value.toString());

        while (itr.hasMoreTokens()) {
            word.set(itr.nextToken());
            if (word.getLength() > 0)
                foo(word);  
            context.write(word, one);
        }
    }

   public static void foo(Text value) {
        value.clear();
   }
}
\end{lstlisting}
\end{minipage}
\caption{Code snippet of simulated Attack \#1 (WordCount)}
\label{fig:wordcount_attack}
\end{figure}

\begin{figure}[htb!]
\center
\begin{minipage}{.45\textwidth}
\begin{lstlisting}[		    
		basicstyle=\fontsize{8}{11}\ttfamily,
        linebackgroundcolor={%
        \ifnum\value{lstnumber}>8
            \ifnum\value{lstnumber}<16
                \color{red!20}
            \fi
        \fi},
]
      13: aload         4
      15: invokevirtual #8          
      18: ifeq          64
      21: aload_0       
      22: getfield      #4                  
      25: aload         4
      27: invokevirtual #9                  
      30: invokevirtual #10                 
      33: aload_0       
      34: getfield      #4                  
      37: invokevirtual #11   
      40: ifle          50
      43: aload_0       
      44: getfield      #4            
      47: invokestatic  #12           
      50: aload_3       
      51: aload_0       
      52: getfield      #4                  
      55: getstatic     #13                 
      58: invokevirtual #14                 
      61: goto          13
      64: return        

  public static void foo(org.apache.hadoop.io.Text);
    Code:
       0: aload_0       
       1: invokevirtual #15          
       4: return       
\end{lstlisting}
\end{minipage}
\caption{Bytecode snippet of WordCount Attack \#1}
\label{fig:wordcount_attack_bytecode}
\end{figure}

\begin{figure}[htb!]
\begin{minipage}{.45\textwidth}
\begin{lstlisting}[
		basicstyle=\fontsize{8}{11}\ttfamily,
		numbers=none
]
...
year        11
year.       2     
years       14
years,      2
yelling     1
yellow,     1
yells.      1
yen         1
yet         29
yet."       1
....

\end{lstlisting}
\end{minipage}
\caption{WordCount final output (normal execution)}
\label{fig:wordcount_output}
\end{figure}

\begin{figure}[htb!]
\begin{minipage}{.45\textwidth}
\begin{lstlisting}[
	    basicstyle=\fontsize{8}{11}\ttfamily,
		numbers=none
]
...
    1
    1
    1
    1
....
\end{lstlisting}
\end{minipage}
\caption{Disrupted Mapper's output}
\label{fig:wordcount_attack_output}
\end{figure}

\begin{figure}[htb!]
\lstset{xleftmargin=0.5cm}
\begin{lstlisting}[
	    basicstyle=\fontsize{8}{11}\ttfamily,
        linebackgroundcolor={%
        \ifnum\value{lstnumber}=2
            \color{green!35}
        \fi
        \ifnum\value{lstnumber}=8
            \color{green!35}
        \fi},
]
...
years,      1
yellow,     1
...
zenana, 1
zenith  1
|       31
        57336
\end{lstlisting}
\caption{Incorrect final result of WordCount job caused by one compromised Mapper}
\label{fig:wordcount_attack_final}
\end{figure}

The attack is detected by control flow analysis and control flow filtering. The control flow trace shows the diverted control flow which is different from the execution paths in CFG of the original program. V-MR detects the diverted control flow and rejects the computation results. 

V-MR was able to detect the attack by control flow analysis and re-execution was not needed. V-MR's verification result is \texttt{REJECT}.

\subsubsection{Attack Scenario 2}
The attacker's code produced incorrect outputs while the control flow trace shows that it followed a valid control flow path.

In the WordCount program, \texttt{context.write(word, one)} method call emits the output, a $<key, value>$ pair, where the parameter 'one' represents one occurrence of a word.
The attacker's code uses '2' instead of '1', with the modified statement as follows.\\

\texttt{private final static IntWritable one = new IntWritable(2);} \\

This attack does not change the control flow of the program but the modification of the constant variable alters the output. 
For example, the attacker's code produces $<"test", 2>$, which is different from the correct output $<"test", 1>$.
The Figure \ref{fig:wordcount_attack2_normal} and Figure \ref{fig:wordcount_attack2_attack} show the difference in the bytecodes between the original code and the attack code. Figure \ref{fig:wordcount_attack2_result} shows the tampered output of Mapper by the attack.

The control flow trace shows the valid execution path but the re-execution can identify that the output is wrong, which is different from the original output.

Control flow analysis was not sufficient for detecting the attack and re-execution was needed. V-MR was able to detect the attack via partial re-execution and the verification result is \texttt{REJECT}.

\begin{figure}[htb!]
\center
\begin{minipage}{.45\textwidth}
\begin{lstlisting}[
	    basicstyle=\fontsize{8}{11}\ttfamily,
        linebackgroundcolor={%
        \ifnum\value{lstnumber}=5
                \color{red!20}
        \fi},
]
static {};
    Code:
       0: new           #14                 
       3: dup           
       4: iconst_1      
       5: invokespecial #15                 
       8: putstatic     #11                 
      11: return 
\end{lstlisting}
\end{minipage}
\caption{Attack \#2: original bytecode snippet}
\label{fig:wordcount_attack2_normal}
\end{figure}

\begin{figure}[htb!]
\center
\hspace*{\parindent}
\begin{minipage}{.45\textwidth}
\begin{lstlisting}[
	    basicstyle=\fontsize{8}{11}\ttfamily,
        linebackgroundcolor={%
        \ifnum\value{lstnumber}=5
                \color{red!20}
        \fi},
]
static {};
    Code:
       0: new           #14                 
       3: dup           
       4: iconst_2      
       5: invokespecial #15                 
       8: putstatic     #11                 
      11: return        
\end{lstlisting}
\end{minipage}
\caption{Attack \#2: attack bytecode snippet}
\label{fig:wordcount_attack2_attack}
\end{figure}
  
\begin{figure}[htb!]
\begin{minipage}{.45\textwidth}
\begin{lstlisting}[
	    basicstyle=\fontsize{8}{11}\ttfamily,
		numbers=none
]
...
years,      2
yellow,     2
...
zenana, 2
zenith  2
\end{lstlisting}
\end{minipage}
\caption{Incorrect intermediate result of WordCount job by Attack \#2}
\label{fig:wordcount_attack2_result}
\end{figure}

The attacker compromises the program without being detected by early stage of the verification. However, the re-execution will detect the output is different from the original output.

\subsubsection{Attack Scenario 3: Collaborating attackers}
The collaborating attackers' code executes on two different workers, which modifies their computation results as the same output with given the same input, using Attack \#2. 

This collaborating attack is detected by partial re-execution as the outputs of two computations by collaborating workers are different from the outputs of the re-executions by V-MR.
While other replication-based verification approaches that use peer comparisons for checking the integrity cannot detect a collaborating attack unless the majority of the workers are honest. However, our approach can detect collaborative attacks.

Control flow analysis cannot detect the attack, while re-execution was able to detect the attack. V-MR's verification result is \texttt{REJECT}.

\begin{table}[t]
\caption {Detection of exploits} 
\label{tab:detection}
\begin{center}
\begin{tabular}{@{}lllcccc@{}}
\toprule
\bf{Exploits}   &  \bf{Description}   & \bf{FP} & \bf{FN}\\      
\midrule
Cheating worker   &   skipping computation         & No    & Yes \\  
Malicious worker &     malicious modification on map function              & No    & No    \\ 
Malicious worker  &   malicious modificaiton on reduce function            &No  & No      \\ 
Collusion  &   collusion attack                     & No      & No      \\ 
\bottomrule
\end{tabular}
\end{center}
\end{table}

\subsection{Experimental Results}

We measured the efficiency of the prototype of V-MR and the overhead of tracing (with instrumentation) compared to a baseline of simple re-execution of the replication-based verification approach for the evaluation of the verifier.
To measure the overhead introduced by trace logging, we performed experiments with several Hadoop applications that are obtained from \cite{HadoopCookbook_2015}.
We use five MapReduce applications: WordCount, InvertedIndex, and three applications for Web application log analysis using the NASA Web Log dataset, for the experiments.
We measured the execution time of the original code and re-execution code.

\subsubsection{Time and Space Overhead for Tracing}
We performed experiments to study the time and space overhead for recording control flow traces of running Hadoop programs on worker nodes during original execution.

Table~\ref{tab:timeOH} shows the time overhead of V-MR with the instrumentation. While we can optimize further to reduce the tracing overhead in the future, the current prototype of V-MR incurs moderate tracing overhead.
The space overhead for tracing is shown in Table~\ref{tab:spaceOH}. Since the verification is performed by a client's on-demand request and the trace logs can be deleted upon the verification is complete, the space overhead for tracing is not a concern in this work. 

\begin{table}[t]
\caption {Time overhead for tracing (Mapper)} 
\label{tab:timeOH}
\centering
\begin{tabular*}{\columnwidth}{llll}
\toprule
\textbf{Program}  & \textbf{\# input records} & \textbf{w/o tracing}  & \textbf{w/ tracing}   \\ 
\midrule
WordCount   &     6982     &    3171       &   3725 (17.4\%)                        \\ 
InvertedIndex &    6982       &   4006             &     4170 (24\%)                     \\ 
WebLog (HitCount)  &   1891715        &      33188         &    64515 (39.6\%)              \\ 
WebLog (Frequency)  &     18617      &     3683          &        3970 (7.2\%)          \\ 
\bottomrule
\end{tabular*}
\end{table}

\begin{table}[h]
\caption {Space overhead of tracing (Mapper)} 
\label{tab:spaceOH}
\begin{center}
\begin{tabular*}{\columnwidth}{llll}
\toprule
\textbf{Program}   &  \textbf{\# input records} &    \textbf{Size (KB)}  \\ 
\midrule
WordCount   &     6982     &    121                        \\ 
InvertedIndex &    6982       &      182                    \\ 
WebLog (HitCount)  &   1891715        &        4867            \\ 
WebLog (Frequency)  &     18617      &      37        \\ 
\bottomrule
\end{tabular*}
\end{center}
\end{table}

\begin{table*}[h]
\caption {Performance comparison (re-execution time)} 
\label{tab:vmr_performance}
\begin{center}
\begin{tabular*}{0.6\textwidth}{@{}llllll@{}}
\toprule
\multirow{2}{3cm}{Program}  &  \multicolumn{2}{c}{Input Size (\# records)} & \multicolumn{2}{c}{Running Time (ms)} & \multirow{2}{4cm}{\textbf{Speedup}}\\
\cline{2-5} & Baseline & V-MR & Baseline & V-MR \\
\midrule
WordCount   &     6982     &    5865                     & 3171  & 2663 & \textbf{1.19}\\
InvertedIndex &    6982       &      5839                & 4006  & 2804 & \textbf{1.43}\\
WebLog (HitCount)  &   1891715        &        1237097   & 33188  & 17589 & \textbf{1.89}\\
WebLog (Frequency)  &     18617      &      18617        & 3683  & 3683 & \textbf{1} \\
\bottomrule
\end{tabular*}
\end{center}
\end{table*}

\subsubsection{Execution time}
We measured the execution time for each Map and Reduce task of full re-execution and partial re-execution to get the idea of speedup by our partial re-execution approach compared to a naive replication-based approach. 

Table~\ref{tab:vmr_performance} shows the experiment results in terms of the speedup by V-MR. Most applications we used for the experimentation show that the execution time is much faster than the naive full execution even with the small size of input data set, while the Weblog application that finds the frequency of Hit count of URLs did not speed up. We intentionally selected this uncommon type of MapReduce application to simulate the case that our partial re-execution approach is not efficient in reducing the re-execution cost. The application does not have any control flow transfer in the map and reduce function as it simply sorts the results produced by WebLog Hitcount application. There is no option to optimize for the selection of data and the instruction elimination for re-execution. 
In this case, while it is very rare, it can be verified by re-executing on randomly chosen instances, which is similar to other replication-based approaches. 
The effect of speedup of re-execution by input data slicing and program slicing is dependent on the ratio of the slice cut on data and program code. 
The Hadoop applications we used for the experiment in this work have relatively small code size (a small number of instructions) in the map and reduce function with small input data set so did not lead to dramatic speedup. However, more complex applications that have much more instructions and larger input data set can benefit more from the effect of the speedup, which remains to be future work.

\section{Conclusion}
\label{sec_conclusion}
This work demonstrated our approach to practical verification of the integrity of outsourced computation via partial re-execution. We investigated the computation result integrity of MapReduce applications as the case study.
Our verification framework, V-MR can detect the tampered MapReduce computations using partial re-execution without modifying the MapReduce framework or application source code. V-MR reduces the re-execution cost by input data slicing and program slicing techniques based on the control flow and data dependency analysis. V-MR is more effective and considerably more efficient than other naive replication-based verification approaches as it can detect attacks against computation integrity with no false positives and few false negatives (shown in Table~\ref{tab:detection}) while it can also speed up the verification via efficient partial re-execution.

\bibliographystyle{IEEEtran}
\bibliography{vmr_paper}

\end{document}